\documentclass[journal]{IEEEtran}
\usepackage[utf8x]{inputenc}

\usepackage[american]{babel}

\ifCLASSINFOpdf
  \usepackage[pdftex]{graphicx}
\else
\fi

\usepackage{amssymb,amsmath,amsfonts,dsfont}
\usepackage{latexsym}
\usepackage{booktabs}
\usepackage{color}

\newcommand{\uM}{{\mathbf u}}
\newcommand{\w}{{\mathbf w}}
\newcommand{\EE}{{\mathbb E}}
\newcommand{\psiM}{{\boldsymbol{\psi}}}
\newcommand{\wo}{{\mathbf{w}}_{\rm o}}

\begin{document}
%
\title{Censoring Diffusion for Harvesting WSNs}

	
	\author{
	\authorblockN{Jesus~Fernandez-Bes, Roc\'{\i}o Arroyo-Valles, Jerónimo Arenas-García, Jesús Cid-Sueiro\\}
	\authorblockA{Signal Theory and Communications Department, Universidad Carlos III de Madrid, Spain\\
	Email: \{jesusfbes,marrval, jarenas, jcid\}@tsc.uc3m.es}
	
	\thanks{The work was partially supported by projects TEC2014-52289-R, PRICAM S2013/ICE-2933 and PRI-PIBIN-2011-1266.}}

%
%



\maketitle

\begin{abstract}

In this paper, we analyze energy-harvesting adaptive diffusion networks for a distributed estimation problem. In order to wisely manage the available energy resources, we propose a scheme where a censoring algorithm is jointly applied over the diffusion strategy. An energy-aware variation of a diffusion algorithm is used, and a new way of measuring the relevance of the estimates in diffusion networks is proposed in order to apply a subsequent censoring mechanism. Simulation results show the potential benefit of integrating censoring schemes in energy-constrained diffusion networks. 

\ 
\end{abstract}

%
\IEEEpeerreviewmaketitle

\section{Introduction}

Diffusion strategies have shown a huge potential in a number of Wireless Sensor Networks (WSN) applications \cite{sayed2013diffusion}: target localization, collaborative spectral sensing \cite{Diff_WSS_Arroyo14}, smart grid monitoring \cite{erol2012suresense}, etc.

Energy harvesting devices, which are able to extract energy from the environment, are a promising technology to overcome the energy limitations of battery-powered sensor nodes \cite{gunduz2014designing}. However, a correct management of the stochastic and scarce harvested energy is still needed. In that direction, censoring schemes are capable of reducing communication processes among nodes (which are among the most energy-demanding tasks \cite{raghunathan2002energy}) without a significant degradation in network performance whenever the information to be censored from nodes is appropriately selected.

We focus on the problem of distributed estimation over adaptive networks where nodes are able to apply censoring policies in order to make an efficient use of the limited energy resources. Sensor nodes are provided with finite batteries but they are able to harvest energy from the environment. Up to our knowledge, there are few works in the literature of diffusion networks that explicitly take the energy expenditure into account. A notable exception is \cite{gharehshiran2013distributed}, where game theory is used to find an activation mechanism in diffusion networks. The algorithm works in two timescales and a utility function that captures the trade-off between the individual contribution of the estimate and the energy expenditure is defined. Another similar idea was presented in \cite{arroyo2013censoring}, where a censoring strategy for the standard Adapt-then-Combine (ATC) algorithm and non-rechargeable WSNs is proposed. A recent work proposes a multihop diffusion strategy under local and network-wide energy constraints in non-rechargeable WSNs \cite{MultihopDiffusion}.

In this paper, we deal with two ideas that do not have been yet put together to get a better management of the constrained energy resources: the application of censoring strategies and energy-harvesting sensors in adaptive diffusion problems. Furthermore, we use a variation of the ATC algorithm \cite{sayed2013diffusion}, named Decoupled Adapt-then-Combine (D-ATC) \cite{fernandez2015decoupled}, because it has been proven to be a robust approach for asynchronous networks. We use the Balanced Transmitter scheme \cite{fernandez2015mdp} as the censoring strategy to prevent the transmission of uninformative data (estimates) and save energy. In this way, there is only one timescale and just a measurement that quantifies the relevance of the information needs to be defined. We propose to use the decrement of the neighborhood estimation error as the importance function. Although this work is just a preliminary step towards the general objective of implementable strategies for efficient energy management in diffusion networks, numerical results already validate the potential of this approach.


The rest of the paper is organized as follows. In Section \ref{sec:signal_energy}, we present the diffusion strategy, and in Section \ref{sec:energy}, the energy model is introduced. The problem of assigning importance to the messages is studied in Section \ref{sec:imp_assingnment}. Later, in Section \ref{sec:review_censoring}, the chosen censoring algorithm is presented and the whole scheme is summarized. Finally, the paper is closed with a section of numerical experiments to evaluate the performance of this technique.

\section{Diffusion strategy}\label{sec:signal_energy}

Consider a network of $N$ nodes connected with a predefined topology. Two nodes are neighbors if they can share information, and we denote $\mathcal{N}_k$ the neighbors of node $k$ including $k$ itself, whereas $\bar{\cal N}_k$ is the neighborhood of node $k$ excluding $k$. Nodes are battery-limited and equipped with an energy-harvesting device. We tackle a distributed estimation problem, where the goal is to estimate vector $\wo(n)$ in a distributed manner in scenarios where energy is scarce. To that aim; and provided that node $k$ has enough battery, at each time step $n \geqslant 0$, each node $k$ has access to local data, $\{d_k(n), \uM_k (n)\}$ (where $d_k(n)$ is a scalar measurement and $\uM_k (n)$ is a regression vector of length $M$). Measured data are related to parameter vector $\wo(n)$ through the linear model
\begin{equation}
d_k(n)=\uM_k^{T}(n)\wo(n)+v_k(n),
\label{eq:dk_model}
\end{equation}
where $v_k(n)$ accounts for the measurement noise, which is assumed to be a realization of a zero-mean white random process with variance $\sigma_{v,k}^2$, and it is independent of the other variables across the network. 
 
To solve the estimation problem, we make use of the Decoupled Adapt-then-Combine (D-ATC) strategy proposed in \cite{fernandez2015decoupled}. D-ATC is a variation of the standard diffusion ATC algorithm \cite{fernandez2015decoupled}, which is more suitable for the problem we are dealing with.
It works as follows. First, each node $k$ adapts a local estimation of parameter $\wo(n)$ at time step $n$, $\psiM_k(n)$, using only local data. Then, it combines the previous local estimation $\psiM_k(n)$ with the estimations obtained from its neighboring nodes in the previous time step,  $\w_\ell(n-1)$ for $\ell\in\bar{\cal N}_k$, via certain combination weights, $c_{\ell k}(n)$ $ \forall \ell, k \in \{1,\cdots,N\}$, which may vary along time and should be properly adjusted. If an NLMS filter is used in the adaptation step \cite{sayed2008adaptive}, the D-ATC algorithm can be written down as
\begin{align}
\psiM_k(n)&=\psiM_k(n-1)+\tilde{\mu}_k(n) \uM_k(n) \xi_k(n), \label{eq:diffusionNLMS}\\
\w_k(n)&= c_{kk}(n) \psiM_k(n) + \sum_{\ell\in\bar{\cal N}_k}c_{\ell k}(n)\w_{\ell}(n-1),\label{eq:diffusion2}
\end{align}
where ${\xi}_k(n)= d_k(n)-\psiM_k^T(n-1) \uM_k(n)$, $\tilde{\mu}_k(n)={\mu_k}/\left[{\delta+\|\uM_k(n)\|^2}\right]$, $\mu_k$ is the step size, and $\delta$ is a regularization factor to prevent division by zero. In addition, the combination weights must satisfy the following conditions:
\begin{equation}
c_{\ell k}(n)\geq 0 \text{, } \sum_{\ell \in \mathcal{N}_k}c_{\ell k}(n)=1, \forall k,
\end{equation}
and $c_{\ell k}(n) = 0 \text{ if } \ell \notin \mathcal{N}_k$.

If a proper energy management in energy-constrained networks is pursued, communication processes should be minimized without compromising the network performance. To achieve that goal, we apply a censoring scheme in the aforementioned D-ATC diffusion strategy based on the idea that some local combined estimation values, $\w_k(n)$, may be not informative enough (i.e., relevant) for neighboring nodes, so their transmission is useless and involves a waste of energy resources. Hence, before the adaptation step at each time step $n$, node $k$ should make a decision (i.e., takes an action $a_k(n)$) whether to transmit the current estimate $\w_k(n)$ to neighboring nodes, $a_k(n)=1$, or censor it, $a_k(n)=0$, according to its energy budget (or energy state), $e_k(n)$, and the estimated relevance of the current estimate, $x_k(n)$. 

\section{Energy model}\label{sec:energy}

To assure long-term, uninterrupted, and self-sustainable operation in battery-limited nodes, we provide sensor nodes with energy-harvesting modules. So, let us first define the energy dynamics of the system. The battery of node $k$ at time slot $n+1$ can be computed as \cite{fernandez2015mdp}
\begin{equation}
e_k(n+1) = \phi_B\left[ e_k(n) - b_k(n) \right],
\end{equation}
where $e_k(n)$ is the available energy in the current time slot $n$, where  $B$ is the battery size, $\phi_B=\max(\min(\cdot,B),0)$ is a clipping function. The (random) energy cost $b_k(n)$ can be split as
\begin{equation}
b_k(n) =  b_{0,k}(n)+ a_k(n) \Delta_k(n) - h_k(n),
\end{equation}
where the involved variables can be explained as follows:

\begin{itemize}
	\item \textbf{${b_{0,k}(n)}$:} Energy consumed by node $k$ during time slot $n$ when sensing new data. It includes also the energy required for the adaptation process. 
	\item \textbf{${\Delta_k(n)}$:} Extra energy consumed by node $k$ at time slot $n$ when transmitting the local estimate to the neighboring nodes. This information is broadcast, so the energy consumption due to this broadcast transmission includes the communication cost with all the neighbors. This is only consumed if the message is not censored.
	\item \textbf{$h_k(n)$:} Energy harvested from the environment. 
\end{itemize}

A node $k$ with empty battery cannot perform any task: measuring data, adapting its estimation or communicating with its neighbors. In such cases, the neighbors of node $k$ assume that its estimate has not changed, i.e., $\w_k(n) = \w_k(n-1)$, in order to adjust their combination weights.

\section{Relevace of transmitted estimates}\label{sec:imp_assingnment}

Following other censoring approaches \cite{arroyo2013censoring, appadwedula2008decentralized}, each node should be able to locally quantify the relevance of the information (i.e., the combined estimate) to decide whether to transmit it or not to neighboring nodes. Thus, less important information can be discarded and energy may be saved. However, it is not trivial to assign an importance value to the information shared in a network. Different importance functions have been defined in the literature for related detection and estimation problems. For instance, in the ATC censoring scheme of \cite{arroyo2013censoring}, the importance function depends on the product of the local combination weight and the distance between measurements. Similarly, the difference of posterior probabilities given the current measurement has been proposed as the importance function in a decentralized detection problem \cite{fernandez2012decentralized}. 

Here, we propose as importance function the decrement of the neighborhood squared estimation error, defined as 
\begin{equation}
x_k(n) = \max\left\{ \frac{1}{N_k}\sum_{j \in \mathcal{N}_k}{J_j}(n) - J_k(n), 0 \right\},
\label{eq:imp_cd_atc}
\end{equation}
where $N_k$ is the cardinality of $\mathcal{N}_k$, and $J_k(n)$ is a local sample-based estimation of the Mean-Squared Error (MSE), which can be computed as 
\begin{equation}
J_k(n) = 	(1-\alpha_x) \cdot J_k(n-1) + \alpha_x \cdot \check{\xi}^2_k(n-1), \label{eq:cost_censoring}
\end{equation}
where $\check{\xi}_k(n)= d_k(n)-\w_k^T(n) \uM_k(n)$, and $\alpha_x \in \left[0,1\right]$ is a smoothing constant. This importance value can be understood as an approximation to the decrease of the Mean-Squared Error in the neighborhood that the combined estimation which  includes that of node $k$, $\w_k(n-1)$, would achieve. When the estimate of node $k$, $\w_k(n)$, is good,  the smoothed squared error $J_k(n)$ will be lower than the average error of the neighbors, $\frac{1}{N_k}\sum_{j \in \mathcal{N}_k}{J_j}(n)$, and therefore, this estimate is important and should be shared.

By using this importance function nodes have to share an additional scalar value, $J_k(n)$. Clearly, this is not a problem because nodes can communicate this value together with $\w_k(n)$, which may have a large amount of coefficients. Some diffusion schemes with adaptive combiners (e.g. \cite{tu2011optimal}) assume a similar increment in the communication cost.

Although this is just a heuristic and a deeper study of othe strategies to assign importance values to estimates, the simulations in Section \ref{sec:sim_censor_datc} will show its good empirical performance.

\section{Censoring algorithm}\label{sec:review_censoring}

The Adaptive Balanced Transmitter (ABT) censoring algorithm proposed in \cite{fernandez2015mdp} is used. Although this scheme is suboptimal for finite-battery size, it is a computationally cheap adaptive censoring algorithm. The basis of the algorithm is the computation of a constant threshold, which balances the consumed and the harvested energy. Considering this threshold value $\tau_k(n)$, node decisions at each time step $n$ can be computed as
\begin{equation}
x_k(n)\overset{a(n)=1}{\underset{a(n)=0}{\gtrless}}\tau_k(n-1),
\label{eq:decision}
\end{equation} 
where threshold $\tau_k(n)$ is computed using a stochastic gradient method:
\begin{align}
	\tau_k(n) &= \tau_k(n-1) \nonumber\\ & + \eta_{k,n} \big[ \rho_{k,n} a_k(n) 	- (1-\rho_{k,n}) (1- a_k(n)) \big],
	\label{Eq.adaptive_balanced_2}
\end{align}
with $\eta_{k,n}$ being a learning stepsize. Scalar value $\rho_{k,n} = \frac{\overline{ b}_{1,k}}{\overline{ b}_{1,k}-\overline{ b}_{0,k}}$, with $\overline{ b}_{1,k}=\EE\{b_k|a_k=1\}$ and  $\overline{ b}_{0,k}=\EE\{b_k|a_k=0\}$, has to be estimated in a sample-based manner.

Table \ref{tab:CD-ATC} summarizes the diffusion scheme together with the censoring algorithm for energy-harvesting WSNs, which is named Censoring D-ATC (CD-ATC). Note that all the steps will consume energy according to the model in \ref{sec:energy}.

\begin{table}[t!]
	\centering
	\begin{normalsize}	
		\begin{tabular} {@{}l@{}}
			\toprule
			\textbf{CD-ATC Scheme} \\
			INPUTS: Initial battery $e_k(0)$ and $\eta$ for all $k$	\\
			\midrule			
			Initialize $\tau_k(n)= 0$, $\bar{b}_{0,k}=0$, $\bar{b}_{1,k}=0$ for all $k$. \\
			\addlinespace	
			At each time step $n$, and for each sensor node $k$:\\							
			\ 1. Sense $\{d_k,\uM_k\}$ and receive estimates from non-censoring\\
			\ \ \ \ \  neighbors: $\{\w_\ell(n-1)\}_{\ell \in\mathcal{N}_k}$. \\
			\ 2. Compute the importance $x_k(n)$ using \eqref{eq:imp_cd_atc}. \\
			\ 3. Decide about transmitting the current estimate $\w_k(n)$ \eqref{eq:decision} \\
			\ 4. Update $\tau_k(n)$ using \eqref{Eq.adaptive_balanced_2}. \\
			\ 5. If $e(n+1)>0$: \\
			\ \ \ \ \ Adapt local estimation $\psiM_k(n)$ using \eqref{eq:diffusionNLMS}.\\			
			\ \ \ \ \ Update combination weights $\boldsymbol{c}_k$ according to \cite{fernandez2015decoupled}.\\
			\ \ \ \ \ Update $J_k(n)$ using \eqref{eq:cost_censoring}.\\
			\ \ \ \ \ Combine received estimations $\{\w_\ell(n-1)\}_{\ell \in\mathcal{N}_k}$\\
			\ \ \ \ \ \ \ \ \ \ with $\psiM_k(n)$ using $\boldsymbol{c}_k$, see \eqref{eq:diffusion2}. \\		
			\ \ \ \ \ Transmit combined estimation $\w_k(n)$ and $J_k(n)$ to the \\
			\ \ \ \ \ \ \ \ \ \ neighbors if $a_k(n)=1$.\\
			\bottomrule		
		\end{tabular}
	\end{normalsize}
	\caption[Censoring D-ATC scheme]{Description of Censoring D-ATC scheme.}
	\label{tab:CD-ATC}	
\end{table}

\section{Simulation Results}\label{sec:sim_censor_datc}

\begin{figure}[t]
	\centering
	\begin{tabular}{cc}
		\includegraphics[width=0.4\linewidth]{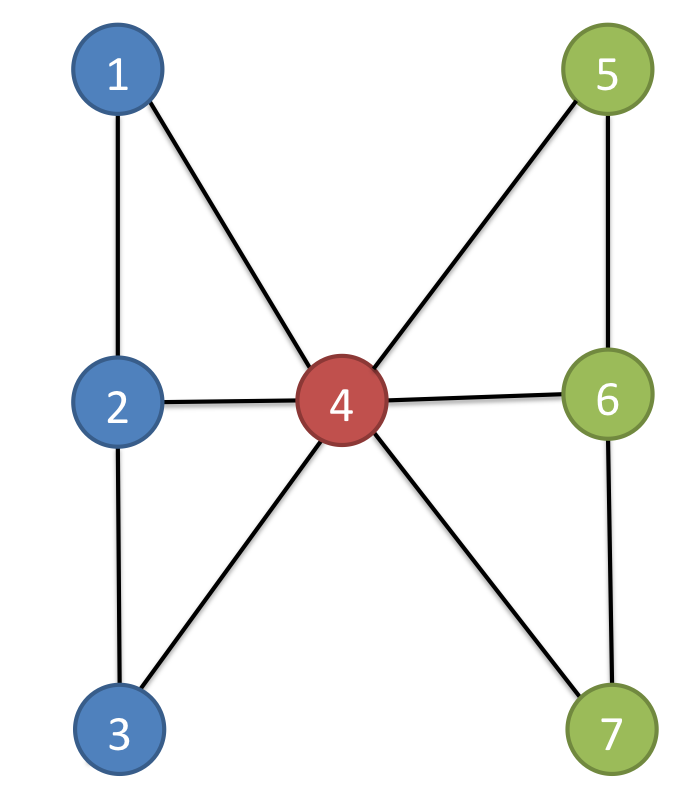}&
		\includegraphics[width=0.5\linewidth]{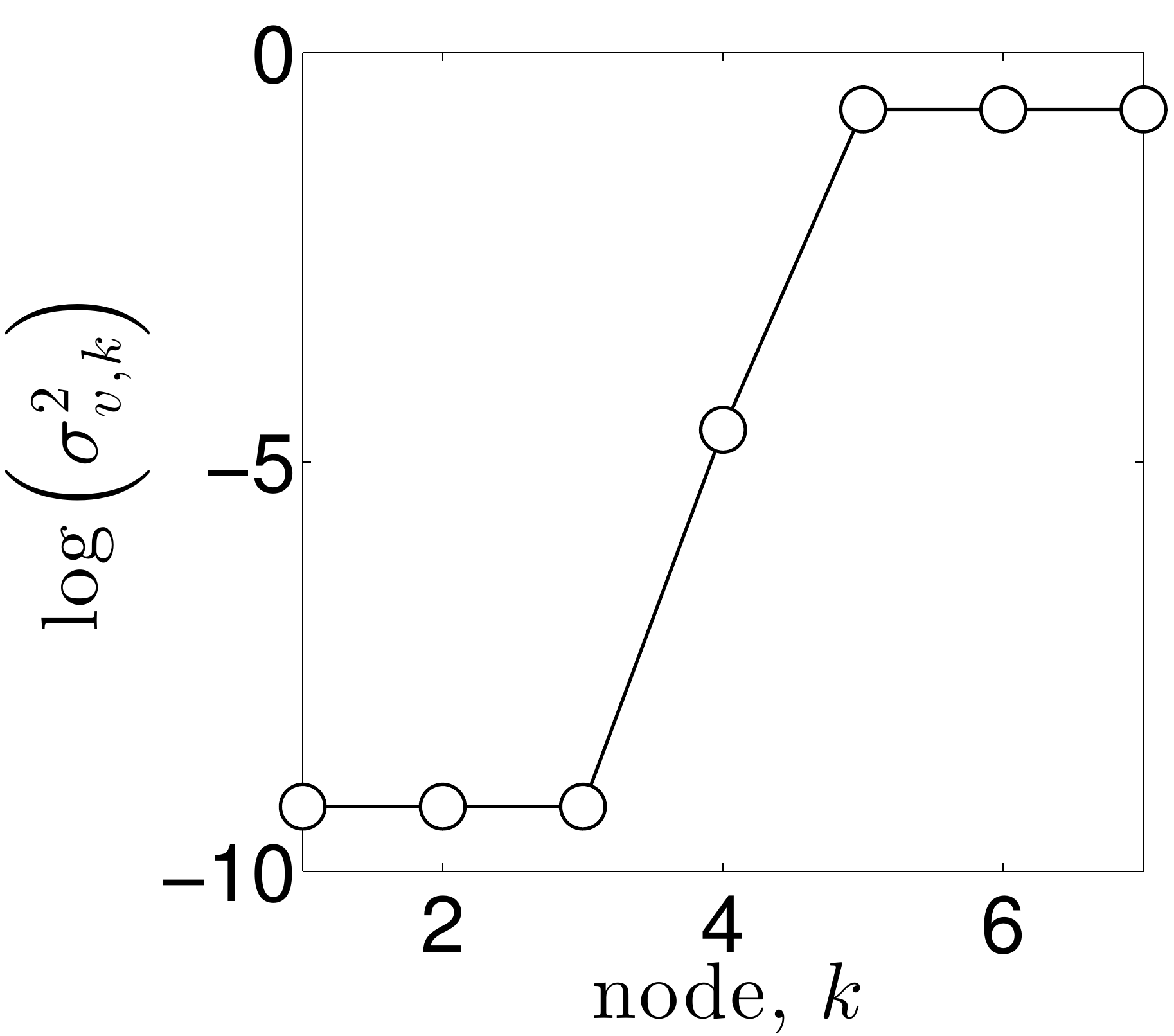}\\
		(a) & (b) \\
	\end{tabular}
	\caption[Network topology and noise variance for C-DATC simulation experiments]{(a) Network topology for the simulation experiments. (b) Noise power $\sigma^2_{v,k}$ at each node in the network (in $\log$-scale).}
	\label{fig:net_cd_atc}
\end{figure}

In this section, we present simulation results to assess the potential of using the proposed CD-ATC algorithm. We consider the network topology of Fig. \ref{fig:net_cd_atc}-(a). All nodes use an NLMS algorithm in the adaptation step, with common step size $\mu=0.1$. The signal power is equal to $1$ for all the nodes and $\wo$ is a 50-taps vector. The only difference among nodes is their noise variances, which is shown in Fig. \ref{fig:net_cd_atc}-(b). In this topology, two subnetworks are connected through node $4$. Nodes $\{1,2,3\}$ are less noisy ($\sigma_{v,\{1,2,3\}}^2=10^{-4}$), while nodes $\{5,6,7\}$ are much noisier ($\sigma_{v,\{5,6,7\}}^2=0.5$), and their steady-state performance is expected to be worse. Node $4$ is a bridge between both subnets and has an intermediate noise variance $\sigma_{v,4}^2=0.01$. Consequently, node $4$ should not be very selective so that the information flows from the left to the right side. Finally, in the combination step we use the \emph{Least-Squares} adaptive combiner rule proposed in \cite{fernandez2015decoupled}.

\begin{figure}[t!]
	\centering
	\begin{tabular}{c}
		\includegraphics[width=0.9\linewidth]{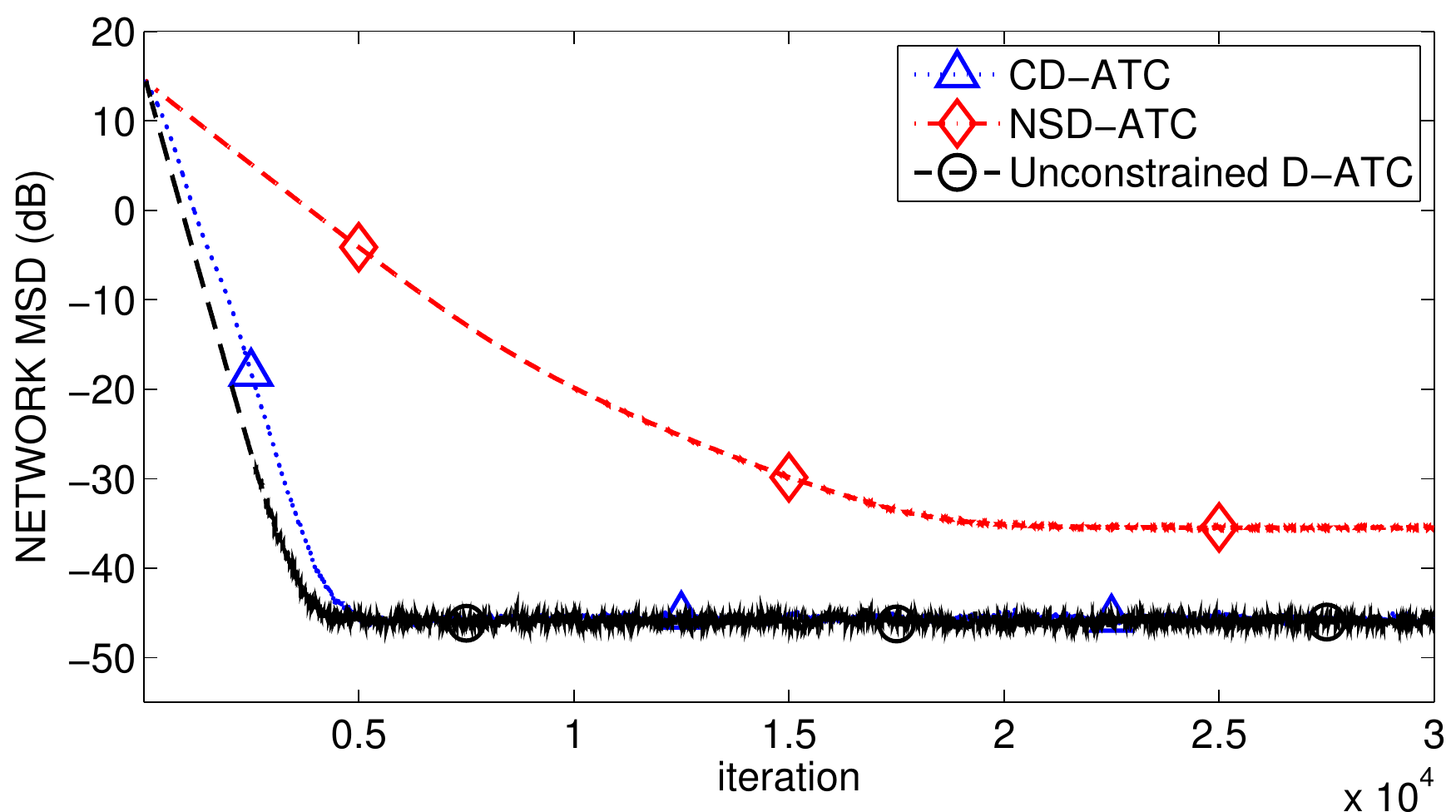}\\
		(a) $p_h=0.4$ \\
		\includegraphics[width=0.9\linewidth]{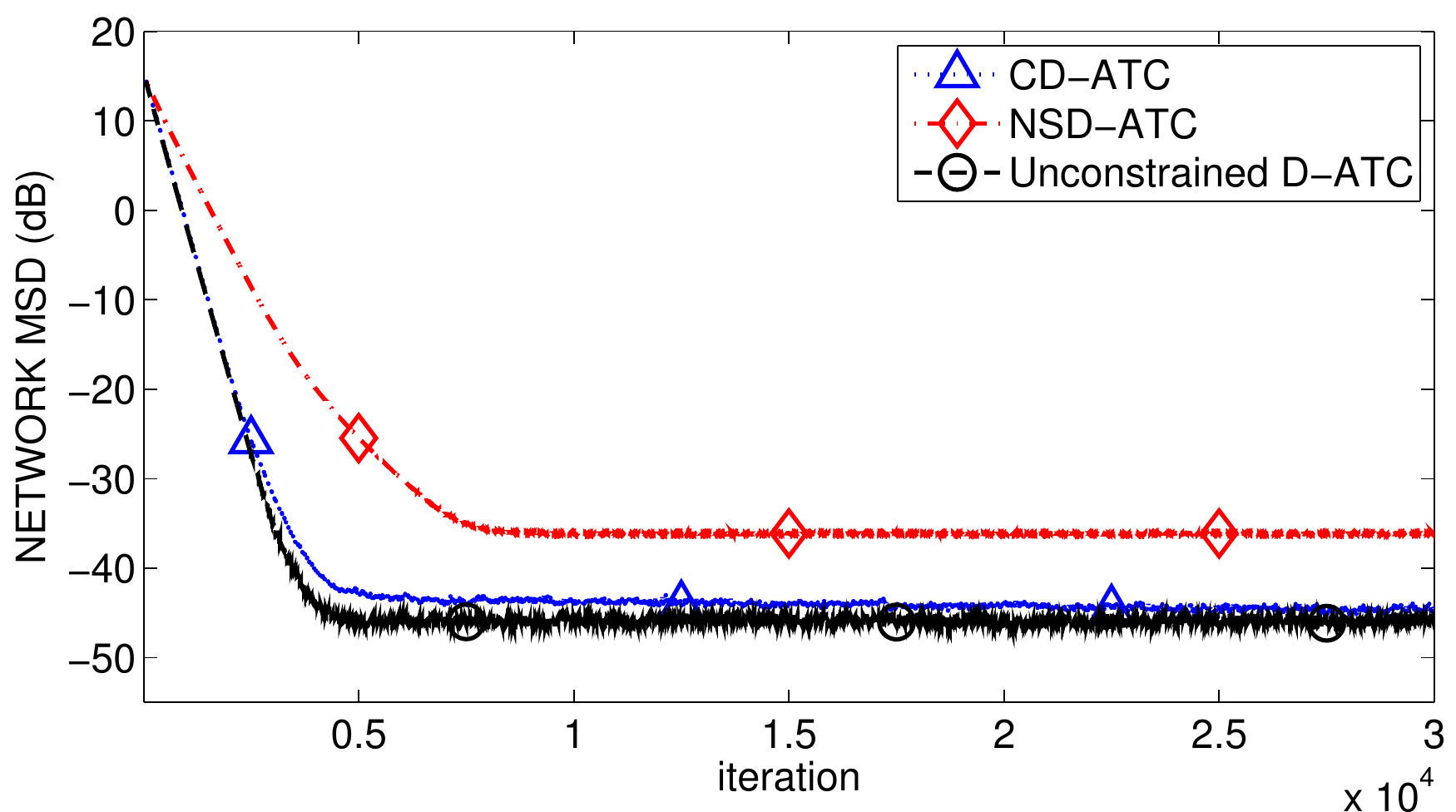}\\
		 (b) $p_h = 0.8$
	\end{tabular}
	\caption[Network MSD for two different harvesting scenarios]{Network MSD performance for two different harvesting scenarios: (a) $p_h=0.4$ and (b) $p_h = 0.8$.}
	\label{fig:nmsd_censor}
\end{figure}

Regarding the energy parameters, all the nodes have the same characteristics: Battery size $B=500$, the sensing cost $b_{0,k}(n)=1$, and the transmission cost $\Delta_k(n)=2$. Finally, nodes randomly refill some energy $h_k(n)$ with probability $p_h$, for which we explore values $0.4$ and $0.8$ to test very different situations. The actual refilled energy is uniformly distributed in the interval $\left[ 2,4\right]$. Fig \ref{fig:nmsd_censor} displays the Network {Mean-Square} Deviation, which is computed as  
\begin{equation}
\text{NMSD}(n) = \frac{1}{N} \sum_{k=1}^N  \EE \left\{ \left[\wo(n) - \w_k(n)\right]^2\right\}\nonumber ,
\end{equation}
for the two different harvesting probabilities $p_h$. The simulated schemes are a non-selective D-ATC, NSD-ATC (i.e., the D-ATC scheme without censoring any information), and the proposed CD-ATC. Note that in both schemes, whenever the battery of a node is depleted, the sensor drops the estimation. Additionally, we also display, as a baseline, the performance of the standard D-ATC in the unconstrained scenario, i.e., infinite amount of energy refill and no censoring.

From both figures, we can conclude that censoring provides an obvious gain both in convergence and steady-state performance. As expected, the gain is larger when the harvesting probability is lower, because in such a case the NSD-ATC battery is more often zero. Even more, the proposed CD-ATC achieves a performance very close to the baseline unconstrained D-ATC in both cases.

\begin{figure}[t]
	\centering
	\begin{tabular}{cc}
		\includegraphics[width=0.9\linewidth]{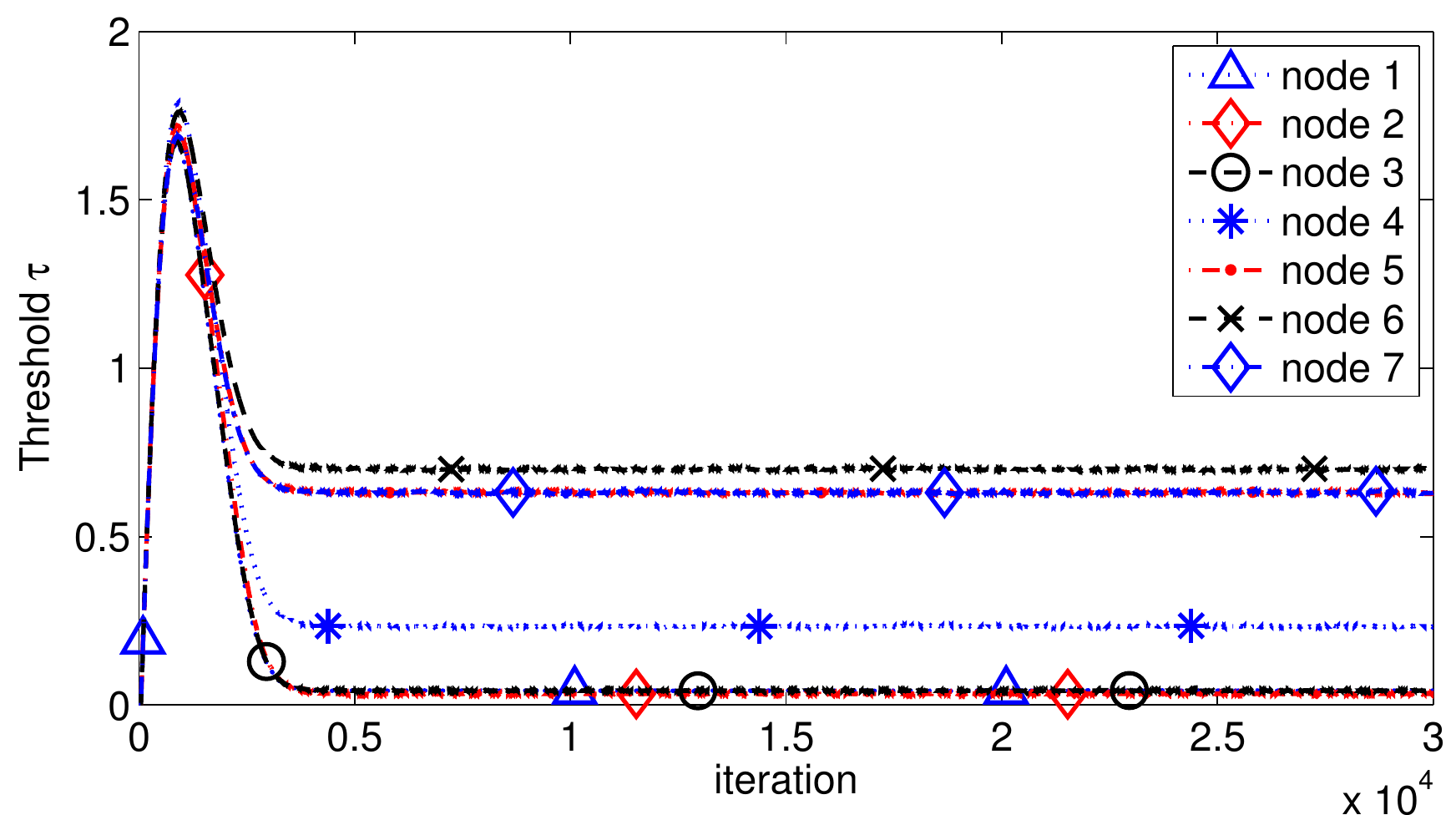}\\
		(a)  $p_h=0.4$   \\
		\includegraphics[width=0.9\linewidth]{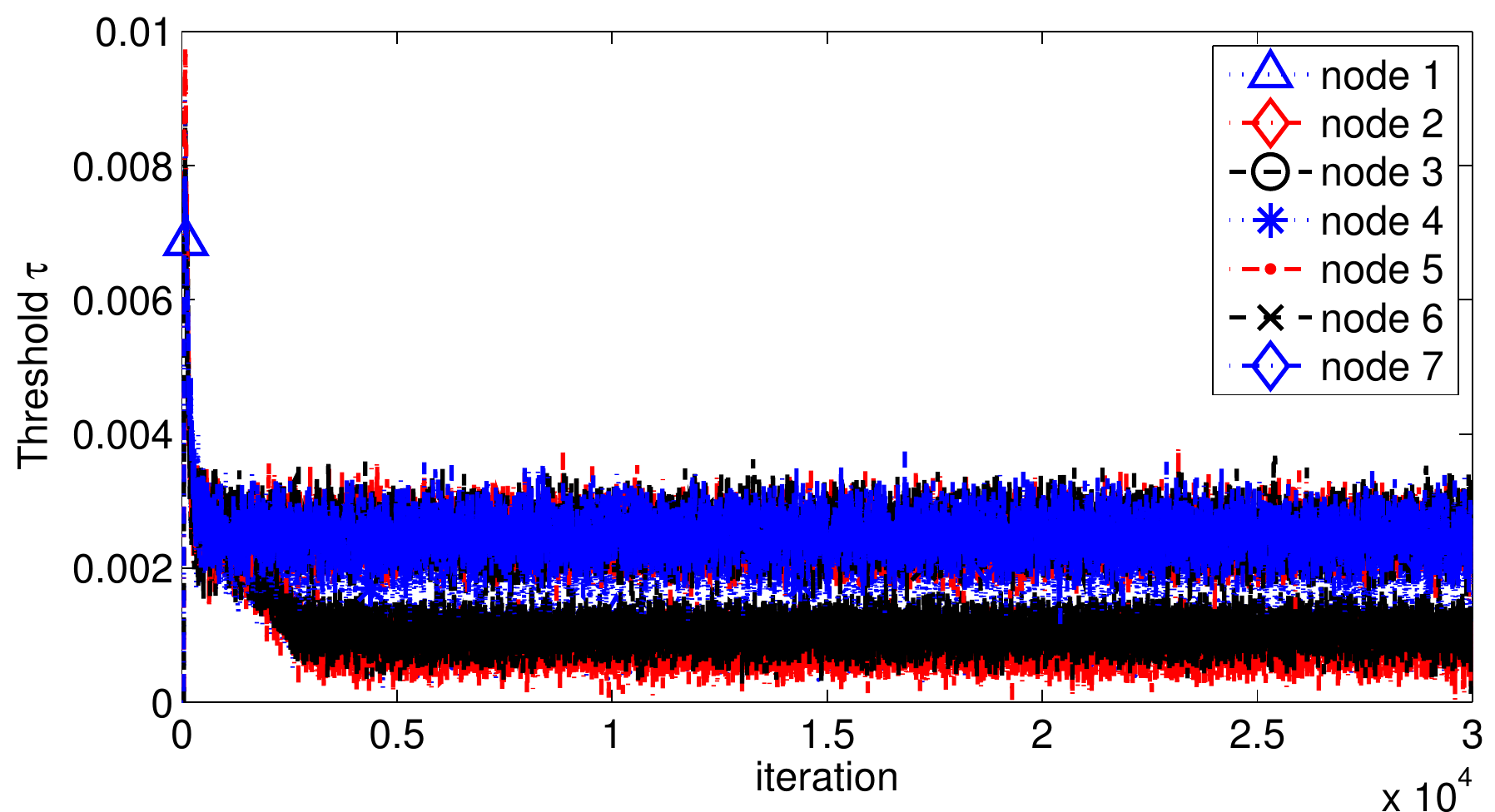}\\
		 (b) $p_h = 0.8$
	\end{tabular}
	\caption[Censoring threshold evolution]{Censoring threshold evolution for all the nodes in the network in two different  harvesting scenarios: (a) $p_h=0.4$ and (b) $p_h = 0.8$.}
	\label{fig:theshold_diffusion}
\end{figure}

In order to understand the behavior of the censoring scheme, we plot in Fig. \ref{fig:theshold_diffusion} the evolution of the thresholds $\tau_k(n)$ for all the nodes. When $p_h = 0.8$ (see Fig. \ref{fig:theshold_diffusion}-(b)), very little censoring ---small threshold values $\tau_k$--- is needed to compensate the energy consumption (nodes refill quite often), and {improve the network performance.} The evolution of $\tau_k(n)$ in Fig. \ref{fig:theshold_diffusion}-(a) is more interesting, where two phases can be observed. In the transient, the thresholds converge to a similar value ---as convergence rate of the nodes does not depend on their noise variance--- and the slight difference among them depends just on the node degree. Then, when nodes are close to reach the steady-state regime, all the thresholds quickly converge to values that mostly depend on the noise variance.

\section{Conclusions}

In this paper, we have proposed an energy-aware censoring diffusion strategy for WSNs whose nodes are equipped with energy-harvesting devices. We have combined the Decoupled Adapt-then-Combine (D-ATC) diffusion algorithm with a balanced censoring scheme. To that aim, we have proposed a function that measures the importance of the computed estimations, which was needed to apply the censoring mechanism. This function approximates the improvement in terms of MSE in a node's neighborhood. Simulated scenarios showed the advantages of using these schemes in energy-constrained environments.

The good performance achieved by the combined scheme suggests that a better design of the importance function or a more involved decision scheme, e.g., random policies, could eventually improve the performance of the standard D-ATC, even in the unconstrained case. In some way, this connects with the design of sparse combination schemes for diffusion networks, which is a topic that, as far as we know, has not been deeply studied in the literature.

\end{document}